%% file: mrejkuba.tex
\documentclass{aa}
\usepackage{graphicx}

\begin{document}
\title{Long Period Variables in NGC~5128: I.\@ 
Catalogue\thanks{Based on observations collected at the European
Southern Observatory, Paranal, Chile, within the Observing Programmes
63.N-0229, 65.N-0164, 67.B-0503, 68.B-0129 and 69.B-0292, and at La Silla  
Observatory, Chile, within the Observing Programme 64.N-0176(B).}
\thanks{ Tables 3-8 and Figures 7-14 together with the complete Sect.~6    
appear only in the electronic version of the article.}}

\author{M. Rejkuba\inst{1,2}
    \and D. Minniti\inst{2}
    \and D.R. Silva\inst{1}}

\offprints{M. Rejkuba}

\institute{European Southern Observatory, Karl-Schwartzschild-Strasse
           2, D-85748 Garching, Germany\\
           E-mail: mrejkuba@eso.org, dsilva@eso.org
  \and Department of Astronomy, P. Universidad Cat\'olica, Casilla
        306, Santiago 22, Chile\\
        E-mail: dante@astro.puc.cl}

\date{Received 05 March 2003 / Accepted 06 May 2003}
\authorrunning{Rejkuba et al.}
\titlerunning{LPV catalogue in Cen A}

\abstract{The first variable star
catalogue in a giant elliptical galaxy NGC~5128 (Centaurus A) is presented.
Using multi-epoch observations with ISAAC at the VLT 
we have detected 1504 red variables in two halo fields covering
$10.46$ arcmin square.
For the variables with at least 10 good measurements, periods and amplitudes 
were determined using Fourier
analysis and non-linear sine-curve fitting algorithms. The final catalogue
contains 1146 long period variables with well established light curve
parameters. Periods, amplitudes, $J_{\rm s}HK_{\rm s}$ photometry as well as
individual $K_{\rm s}$-band magnitudes are provided 
for all the variables.
The distribution of amplitudes ranges from 0.3 to
a few magnitudes in the $K_{\rm s}$-band, with a median value around 0.7~mag.
The amplitudes, mean magnitudes and periods indicate that the
majority of variables belong to the class of long period variables
with semiregular and Mira variables. Exhaustive simulations
were performed in order to assess the completeness of our catalogue and the
accuracy of the derived periods. 
}

\maketitle

%
%
\section{Introduction}

In a stellar population older than few hundred Myr the near-IR light
is dominated by red giants. Of these, first ascent giants
are the brightest among the metal-poor stars older than 1-2 Gyr.
In composite stellar populations, like those found in galaxies, the first
ascent giants are mixed with asymptotic giant branch (AGB)
stars. The red giants fainter than the tip of the first ascent giant
branch are usually referred to as red giant branch (RGB) stars. In the
following RGB is used to denote the first ascent giants, although it
should be kept in mind that it is not possible to
separate RGB and AGB stars that are fainter than the RGB tip.

In intermediate-age populations
($\sim 1 - 5$~Gyr old) numerous bright asymptotic giant
branch (AGB) stars are located above the tip of the RGB.
However, among old populations like
Galactic globular clusters with ${\rm [Fe/H]}\ga -1.0$~dex
and in the Galactic bulge, bright stars have also been
detected above the tip of the RGB (Frogel \&
Elias~\cite{frogel&elias88}, Guarnieri et al.~\cite{guarnieri+97}),
implying the presence of bright AGB stars
in metal-rich and old populations. Essentially all of the bright
giants above the RGB tip in globular clusters seem to be long period
variables (LPVs; Frogel \& Elias~\cite{frogel&elias88}, Frogel
\& Whitelock~\cite{frogel&whitelock98}).
The frequency of LPVs in old metal-rich globular clusters of the MW
and in the Bulge has been studied by
Frogel \& Whitelock~(\cite{frogel&whitelock98}).
Old populations of lower metallicity are known not to
have AGB stars brighter than the RGB tip. The presence
or absence of these bright giants located above the tip of the first
ascent giant branch has important implications
for the magnitude of the surface brightness fluctuations specifically
in the near-IR (e.g. Mei et al.~\cite{mei+01}, 
Liu et al.~\cite{liu+02}), a method used
to determine distances to galaxies that are too distant to
have their stellar content resolved,
but that still present fluctuations due to the underlying light
of giant stars (Tonry \& Schneider \cite{tonry&schneider88}).

The properties of LPVs have been reviewed by Habing~(\cite{habing96}).
They are thermally pulsing asymptotic giant branch (TP-AGB) stars with
main sequence masses between 1 and 6 M$_\odot$. They present variability
with periods of 80 days or longer, and often the longest period variables
show variable or multiple periods. Two main classes of LPVs are
Mira variables (Miras) and semiregular variables (SRs). SRs usually have
smaller amplitudes as well as shorter periods and more irregular light curves 
than
Miras. SRs are sometimes subdivided into subclasses (SRa, SRb)
depending on the regularity and multiplicity of their periods and
shape of their light curves.
The separation between Miras and SRs is not always 
clear (e.g.\ Kerschbaum \& Hron \cite{kerschbaum&hron92,kerschbaum&hron94}).
The classical definition requires that Miras have $V$-band
amplitudes larger than 2.5~mag and regular periods in the range of
80--1000 days (GCVS Kholopov~\cite{gcvs}).
Mean $K$-band amplitudes of Miras are $\sim0.6$~mag
(e.g. Feast et al.~\cite{feast+82}, Wood et al.~\cite{wood+83}).

In the gE NGC~5128 Soria et al.~(\cite{soria+96}),
based on $VI$ HST CMD, and Marleau
et al.~(\cite{marleau+00}), based on $JH$ NICMOS data, suggested a presence of
up to 10\% bright AGB stars belonging to an intermediate-age population.
Harris et al.~(\cite{harris+99,harris&harris00}) on the contrary
did not find bright AGB stars in their $VI$ CMDs of two halo fields
in NGC~5128. However, $V$ and $I$ bands are not
very sensitive indicators of these cool giants and thus some of them might
have been confused with the RGB tip or foreground stars 
and few stellar blends or stars
with larger photometric errors. 

Our group has obtained VLT images with FORS1
and ISAAC in the halo of NGC~5128 in order to
study the bright giants in the halo of the galaxy 
(Rejkuba et al.~\cite{rejkuba+01}). We have observed two halo fields in
the $V$ and $K_{\rm s}$-bands finding
a large number of bright AGB stars, extending up
to bolometric magnitude of --5. 
Field~1 coincides with the prominent north-eastern diffuse
stellar shell (Malin et al.\ \cite{malin+83}, Rejkuba et
al.~\cite{rejkuba+01}). Crossing it in 
diagonal there is a chain of young stars (Mould et al.~\cite{mould+00},
Fassett \& Graham~\cite{fg00},
Rejkuba et al.~\cite{rejkuba+02}). Field~2 is located away from the known
stellar shells and dust bands,  $\sim 9\arcmin$ south from the galactic
nucleus. It coincides with the field observed with HST in $V$ and $I$-bands
with WFPC2 by Soria et al.~(\cite{soria+96}) and in $J$ and $H$-bands with
NICMOS by Marleau et al.~(\cite{marleau+00}). Part of the VLT ISAAC
$K_{\rm s}$-band data analyzed here were already described in
Rejkuba et al.~(\cite{rejkuba+01}). 
Already with the first few epochs of $K_{\rm s}$-band imaging it was
obvious that variable stars were present among the bright red giants. The
preliminary results of the multi-epoch imaging in $K_{\rm s}$-band 
were presented by
Rejkuba (\cite{rejkubaPhd}).

We present here the full near-IR data-set containing multi-epoch 
$K_{\rm s}$-band
imaging and single-epoch data in 
the $J_{\rm s}$- and $H$-bands obtained over the period of
3 years with ISAAC at the VLT. 
In this paper the data reduction, the photometry,
and the catalogue of the variable stars are presented. In the following
papers these data are used to investigate the variability 
characteristics of the population of stars found above the RGB tip,
and to derive the distance to NGC 5128
using the Mira period-luminosity relation in the $K$-band. 
Multicolor information will further be used to
investigate the chemical composition of these
variables, and to put constraints to their ages. 
In the next section we briefly present the
data and the reduction procedures. Section~\ref{photometry} 
contains the photometry.
The technique used to detect the
variable stars is described next. The light curves and periods of the
long period variables are derived in Sect.~\ref{LPVperiods}. 
A detailed description of the 
completeness simulations is given only in the electronic version of 
the article where also the complete catalogue of all the variable stars
can be found. Sect.~\ref{compl_cont_section} presents the main results of 
the our simulations, while the last section summarizes the results.

%
%

 \section {The data}
\label{data}

\subsection{The observations}

We have obtained a total of 
20 epochs of $K_{\rm s}$-band photometry in Field~1 and 24 epochs in 
Field~2 between April 1999 and July 2002. 
Most of our data were obtained using the ISAAC near-IR imaging 
spectrometer at the ESO Paranal UT1 Antu 8.2m telescope. 
These data were obtained in Service Mode. One Field~2 epoch comes 
from data obtained in Visitor Mode at the ESO La Silla NTT 3.5m telescope
equipped with SOFI near-IR imaging spectrometer.

The instrument setup for the observations was the following: 
all but 2 epochs for
each field were observed with the short wavelength arm of ISAAC, which is
equipped with a $1024 \times 1024$ pixel Hawaii Rockwell array with 
a pixel scale
of $0\farcs148$. The last two epochs of 
the $K_{\rm s}$-band series for both fields were
observed with the long wavelength arm of ISAAC with 
a $1024 \times 1024$ InSb Aladdin array from Santa Barbara
Research Center. 
The pixel size of the Aladdin array with $0\farcs1478$ is almost identical
to that of Hawaii detector.

\begin{table*}
\centering
  \caption[]
	{Near-IR multi-epoch photometry observing log. On the left Field~1
	observations, and on the right Field~2 observations are described.
MJD is JD--2400000, Exp is the exposure time in minutes and X is the mean
Airmass during the observing sequence.}
    \label{obslog_isaac}
      \begin{tabular}{ccccl|ccccl}
        \hline \hline
MJD  & Exp & X&Seeing    & \multicolumn{1}{c|}{filter}  
& MJD& Exp & X&Seeing    & \multicolumn{1}{c}{filter}  \\
     & min &  &$\arcsec$ & epoch             
&    & min &  &$\arcsec$ & epoch       \\
\hline 
\hline
51277.1&60&  1.15 &0.41&1Ks01&51277.2&60&   1.05 &0.37&2Ks01\\ 
51306.1&60&  1.05 &0.40&1Ks02&51306.1&60&   1.10 &0.38&2Ks02\\ 
51327.1&60&  1.14 &0.52&1Ks03&51328.0&60&   1.06 &0.44&2Ks03\\ 
51650.1&65&  1.09 &0.36&1Ks04&51594.4&45&   1.04 &0.55&2Ks04$^1$\\  
51675.2&65&  1.22 &0.43&1Ks05&51650.3&65&   1.09 &0.40&2Ks05\\ 
51702.9&65&  1.13 &0.61&1Ks06&51677.2&51&   1.22 &0.52&2Ks06\\
51734.0&65&  1.15 &0.31&1Ks07&51684.2&20&   1.22 &0.44&2Ks07\\
52037.1&60&  1.05 &0.64&1Ks08&51703.0&41&   1.13 &0.58&2Ks08\\
52060.1&60&  1.08 &0.46&1Ks09&51705.0&30&   1.13 &0.41&2Ks09\\
52096.0&60&  1.05 &0.33&1Ks10&51734.1&65&   1.15 &0.40&2Ks10\\ 
52299.3&56&  1.44 &0.55&1Ks11&52037.2&60&   1.10 &0.59&2Ks11\\ 
52309.3&56&  1.27 &0.49&1Ks12&52060.1&60&   1.06 &0.50&2Ks12\\ 
52322.2&56&  1.38 &0.47&1Ks13&52096.1&60&   1.12 &0.38&2Ks13\\ 
52348.3&56&  1.06 &0.47&1Ks14&52300.3&56&   1.30 &0.55&2Ks14\\ 
52360.1&56& 1.17  &0.41&1Ks15&52315.3&56&   1.10 &0.46&2Ks15\\
52379.2&56& 1.06  &0.50&1Ks16&52332.3&56&   1.13 &0.36&2Ks16\\
52403.2&56& 1.27  &0.38&1Ks17&52348.3&54&   1.10 &0.37&2Ks17\\
52422.0&56& 1.10  &0.43&1Ks18&52360.2&56&   1.06 &0.36&2Ks18\\
52444.0&56& 1.11  &0.53&1Ks19$^2$&52379.1&56&   1.08 &0.50&2Ks19\\
52474.0&56& 1.07  &0.49&1Ks20$^2$&52398.1&35&   1.06 &0.38&2Ks20\\
52385.1&70& 1.08  &0.38&1Js  &52405.0&56&   1.30 &0.58&2Ks21\\
52411.1&56& 1.09  &0.47&1H   &52422.1&56&   1.06 &0.49&2Ks22\\
       &  &       &    &     &52444.0&56&   1.06 &0.62&2Ks23$^2$\\
       &  &       &    &     &52474.0&56&   1.16 &0.59&2Ks24$^2$\\
       &  &       &    &     &52385.2&70&   1.06 &0.48&2Js\\
       &  &       &    &     &52369.3&56&   1.17 &0.50&2H \\
\hline
$(1)$&\multicolumn{7}{l}{Observed with SOFI@NTT}\\
$(2)$&\multicolumn{7}{l}{Observed with LW arm of ISAAC@VLT}\\
        \end{tabular}
\end{table*}

During the service observations of 3 epochs of Field~2, 
the integration was interrupted before the end of the sequence  
due to changing weather conditions. Additional observations for these
epochs were taken a few days after the original sequences. 
Thus we reduced them as separate epochs, yielding 24
data points for Field~2, including the epoch observed with SOFI at NTT, 
and 20 epochs
for Field~1. Additional $J_{\rm s}$- and 
$H$-band single epoch observations were
obtained in April and May 2002 with ISAAC, also in service mode.
$J_{\rm s}$ and $H$-band observations were secured with a
short wavelength arm of ISAAC. 
The $J_{\rm s}$ filter was preferred over the $J$ filter, due
to the presence of red leaks in the $K$-band of the latter. 
From here on ``$J$-band'' is used to denote observations taken 
with $J_{\rm s}$ filter and ``$K$-band'' for $K_{\rm s}$ observations. 
The observation taken with $K_{\rm s}$ filter with SOFI at NTT
is described in Rejkuba (\cite{rejkuba01}).

The summary of the observations is given in
Table~\ref{obslog_isaac}, where on the left we describe the Field~1
observations and on the right the Field~2 observations. For each field, the
first column is the Julian date of the observation 
in the form $\mathrm{MJD} =
\mathrm{JD} - 2400000$, in the second column the exposure time in minutes is
given, the next is airmass, then seeing in arcseconds, and in the last field
the filter and the sequential number of the $K$-band epoch are presented.
 
\subsection{Data reduction}

Data reduction followed the procedure described in
Rejkuba et al. (\cite{rejkuba+01}). We first correct for the
electronic ghost using the ESO supplied routine in the ECLIPSE package.
Then we subtract the dark and divide with the sky flats provided
from the service observing standard calibrations. The DIMSUM package
(Stanford et al.~\cite{dimsum}) in
IRAF was used to subtract the sky in a double sky-subtraction run,
the second time masking the objects detected in the registered and
combined frames after the first sky subtraction. The
double sky subtraction procedure is
particularly important for these crowded images. We noticed that in a
single sky subtraction pass, bright regions were often over-subtracted
producing dark halos around the bright stars. Finally the sky-subtracted
images were aligned with IMALIGN and all the images taken in a single
jittered sequence were combined with IMCOMBINE task in IRAF.

\section{The photometry}
\label{photometry}

The PSF fitting photometry was done for each
single epoch image individually. First we used DAOPHOT{\sc ~II}
\textnormal{and} ALLSTAR
(Stetson~\cite{stetson87}) to create a PSF for each frame and to
define the coordinate transformations between the frames. The complete
star list was created from the median combined image that had only the
best seeing epochs (i.e. those with FWHM stellar profiles measured $<3$~pix). 
These included epochs 1Ks01, 1Ks02, 1Ks04, 1Ks05, 1Ks07, 1Ks10, 1Ks15,
1Ks17, 1Ks18 and 1Js for Field~1, and
epochs 2Ks01, 2Ks02, 2Ks05, 2Ks09, 2Ks10, 2Ks13, 2Ks16, 2Ks17 and 
2Ks18
for Field~2. The total exposure
time of these median combined images are 9.4 and 8.4 hours.
We found that better results were obtained in this way instead
of combining all the images. PSF fitting photometry using this star
list and coordinate transformations was performed simultaneously
on all images of a single field with the 
ALLFRAME programme (Stetson~\cite{stetson94}). The
final photometric catalogue contains 13111 stars
in Field~1 and 16435 stars in Field~2, that have been detected on
at least 3 $K$-band frames and in $J$- and $H$-band images.  
The areas covered with at least three 
pointings in $K$-band are  $2\farcm28 \times
2\farcm30$ and $2\farcm25 \times 2\farcm31$, and the total exposure times in
$K$-band are 
19.67 and 21.17 hours, for Fields 1 and 2, respectively. {\it These
are the deepest near-IR images obtained so far in an external galaxy halo.}

\subsection{The photometric calibration}

\begin{figure}
\centering
\includegraphics[width=6.55cm,angle=270]{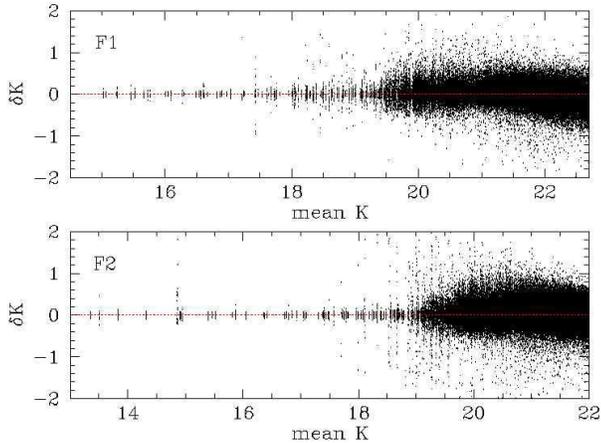}
  \caption[]
	{Photometric accuracy for the $K$-band data. Plotted are 
	differences of individual epoch measurements with 
	respect to their mean magnitudes. Only individual 
	measurements with ALLFRAME $\sigma$ error smaller than 
	0.2 are plotted. At the	bright end the limiting accuracy of 
	$\pm0.05$ mag can be seen. The variables are
	located where an excessive scatter with respect to the mean is
	present with both positive and negative excursions.}
  \label{plot_magdelta}
\end{figure}  

Not all the epochs were observed in photometric conditions. The $K$-band
photometry was brought to the system of one of the photometric nights that
had excellent seeing: epochs 1Ks07 and 2Ks10 for Fields 1 and
2, respectively, both observed on July 8, 2000. The zero point of
$24.257\pm0.042$ for that night has been derived from the observations of
Persson et al.~(\cite{persson+98}) 
standards supplied by ESO service observing (Rejkuba
et al.~\cite{rejkuba+01}).
$J$- and $H$-band 
photometric zero points were derived using the observations of
Persson standard stars obtained during the same night as the NGC~5128
observations  in the corresponding filter (see Table~\ref{obslog_isaac}) 
and their values are $24.737\pm0.039$  and $24.626\pm0.033$. 
For all three filters
extinction coefficients measured by ESO observatory staff and
reported on ISAAC web pages were assumed 
(0.06 for $K_{\rm s}$ and $J_{\rm s}$-band and 0.05 for $H$-band). 

The quality of the photometry can be assessed from 
Figure~\ref{plot_magdelta} where we plot differences of individual epoch
measurements vs. mean magnitudes. The mean magnitudes were
calculated weighting the individual measurements
with their photometric errors calculated by ALLFRAME:
\begin{equation}
\overline{K} = \frac{\sum\limits_{i=1}^n
\frac{K_i}{\sigma^2_i}}{\sum\limits_{i=1}^n \frac{1}{\sigma^2_i}}
\label{meanK.eqn}
\end{equation}
Only the stars with individual ALLFRAME $\sigma$ uncertainty measurements
smaller than 0.2 mag are
plotted. The limiting accuracy of our photometry at the bright end is $\pm
0.05$ mag, getting worse at faintest magnitudes. The region around K=20
has more scattered measurements than the mean. This is where most variable
stars are found. The mean magnitude calculated with the above equation will,
however, be overestimated (too bright) for variable stars with large amplitudes
because brighter phases will typically get higher weights. 

%
%

\section{Search for variable stars}
\label{varsearch}

\begin{figure}
\centering
\includegraphics[width=6.55cm,angle=270]{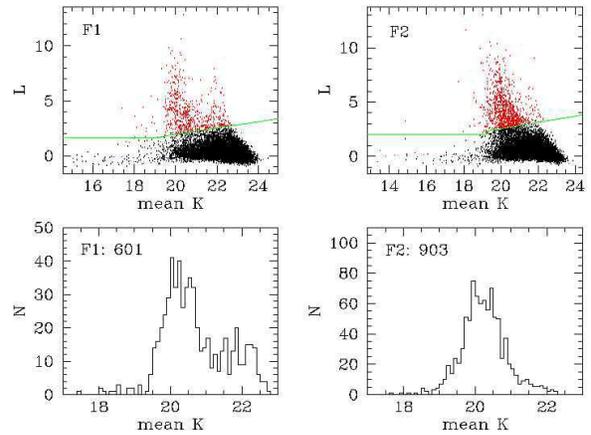}
  \caption[]
	{Variability index L vs. magnitude is plotted in the upper two panels
	for fields 1 (left) and 2 (right). 
	Variable stars have higher index L values than non-variables and are 
	found above the solid line.
	In the lower panels the magnitude
	distribution of variable stars is shown. The number of variables
	with at least 6 data points is also given.}
  \label{mag_varindex}
\end{figure}  

Variable stars were identified using a procedure similar
to the one described by Welch \& Stetson~(\cite{welch&stetson93}) 
and Stetson~(\cite{stetson96}). First, 
we selected all the stars with mean of all photometric errors given by
ALLFRAME, as measured on each individual epoch frame, 
smaller than 0.2~mag. We then required for each star to be detected
on more than 5 frames 
and constructed variability indexes J, K and L, following the prescriptions
by Stetson~(\cite{stetson96}; equations~1--3). 

In particular the index J is used for single observations assuming
$P_i=\delta^2_i - 1$, where $\delta_i$ is a normalized residual of the
measurement from the mean magnitude.
The mean magnitude is calculated using a weighting with
inverse square of the measurement error (weight $w_i$) according to 
equation~\ref{meanK.eqn}. 

The index J is a robust measure of the external repeatability
relative to the internal precision. For single epoch observations
containing only random noise its value tends to zero. 
For a physical variable it is a positive number. 
The index K is a robust measure of the kurtosis of the magnitude
histogram and its value is fixed by the shape of the light curve, with
K=0.9 for a pure sinusoid and K=0.798 for a Gaussian magnitude distribution,
a limit approached when random measurement errors dominate over physical
variation.
Including the information about the variability and the shape of the
light curve, Stetson defined the final variability index L as:
\begin{equation}
L= \left( \frac{JK}{0.798} \right) \left( \frac{\sum w_i}{w_{all}} \right)
\end{equation}
The second term in this expression assigns an additional weight to
stars with the highest number of measurements. 

\begin{figure}
\centering
\includegraphics[width=6.55cm,angle=270]{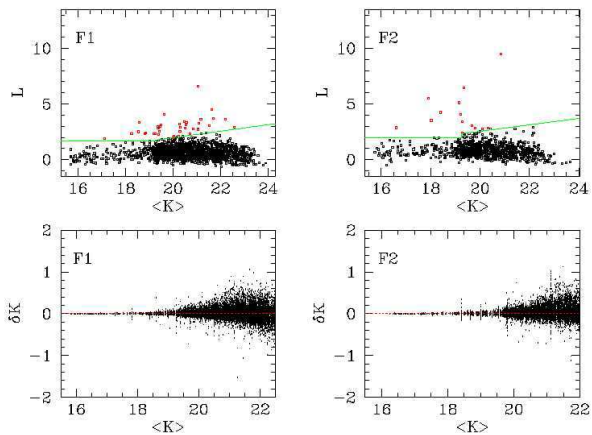}
  \caption[]
	{Variability index L vs. magnitude for simulated non-variable stars,
	i.e. those with light curve constant in time, 
	is plotted in the upper two panels
	for fields 1 (left) and 2 (right).
	This diagram is used to decide the detection limit for the variable
	stars. In the lower panels scatter of the individual magnitude
	measurements around the mean magnitude is shown. }
  \label{magNLaddconst}
\end{figure}  

Figure~\ref{mag_varindex} is a plot of the variability index L for all the
stars in Field~1 and 2 in the
left and right panels, respectively. The stars above the 
solid line (upper panels), plotted
with slightly larger symbols, were searched for periodic variations. 
The limits that
separate variable from non-variable sources were
determined in the following way: in
completeness simulations stars with constant magnitude light curves were
added to the frames and their photometry and variability indexes
were measured in identical way as for the programme stars. A similar plot
of the variability index $vs.$ magnitude was used to decide on the
separation line between the variable and non-variable stars
(Figure~\ref{magNLaddconst}). With the adopted variable star detection
limits (solid lines in the upper panels) 
less than 2\% of sources from these simulations
are found above the lines. Note however, that at least some of these may
have their light curves contaminated by a neighbouring variable star. 
Simulations necessary to evaluate the completeness of the
variable star search are described below. 

With such selection criteria 
601 stars in Field~1 and 903 stars in Field~2 are found to be variable.
Of these 536 and 878 had at least 10 measurements with individual
errors smaller than 0.5 mag, and 
light curves were constructed for them. 

\section{Light curves and periods of LPVs}
\label{LPVperiods}

Fourier analysis of the $K$-band light curves was used to
search for the periodic signal in the data in the range of $30<P<1700$ days. 
Only individual measurements with
ALLFRAME uncertainty smaller than 0.5~mag were used. 
First, an initial guess of the period was obtained using a routine
that calculates spectral power as a function of angular
frequency ($\omega=2\pi f$; Lomb~\cite{lomb}):
\begin{eqnarray}
P_N(\omega)& =& 
\frac{[\sum_j (K_j - \overline{K}) \cos \omega (t_j - \tau) ]^2}
{2 \sigma^2 \sum_j \cos^2 \omega (t_j-\tau)}  \nonumber \\
           & +  &
\frac{[\sum_j (K_j - \overline{K}) \sin \omega (t_j-\tau)]^2}
{2 \sigma^2 \sum_j \sin^2 \omega (t_j-\tau)} 
\end{eqnarray}
where $K_j$ and $\overline{K}$ are the individual and mean
magnitudes, $\sigma$ is the variance, 
$t_j$ is the Julian Date (JD) of the measurement, and $\tau$ is defined as:
\begin{equation}
\tan (2\omega \tau) = \frac{\sum_j \sin 2 \omega t_j}{\sum_j \cos 2 \omega
t_j}
\end{equation}
The period obtained from the frequency with largest power corresponds
to the most probable sinusoidal component. It was further improved
with a non-linear least-square fitting of the function
\begin{equation}
K(t) = A \cos \left( 2 \pi \frac{(t - t_0)}{P} \right) +
B \sin \left( 2 \pi \frac{(t - t_0)}{P} \right) + K_0
\end{equation}
From this, the best fitting period (P), amplitude 
($2\times \sqrt{A^2+B^2}$), mean
magnitude ($K_0$), and phase ($t_0$) were obtained.

\begin{figure*}
\centering
\includegraphics[width=6.6cm,angle=270]{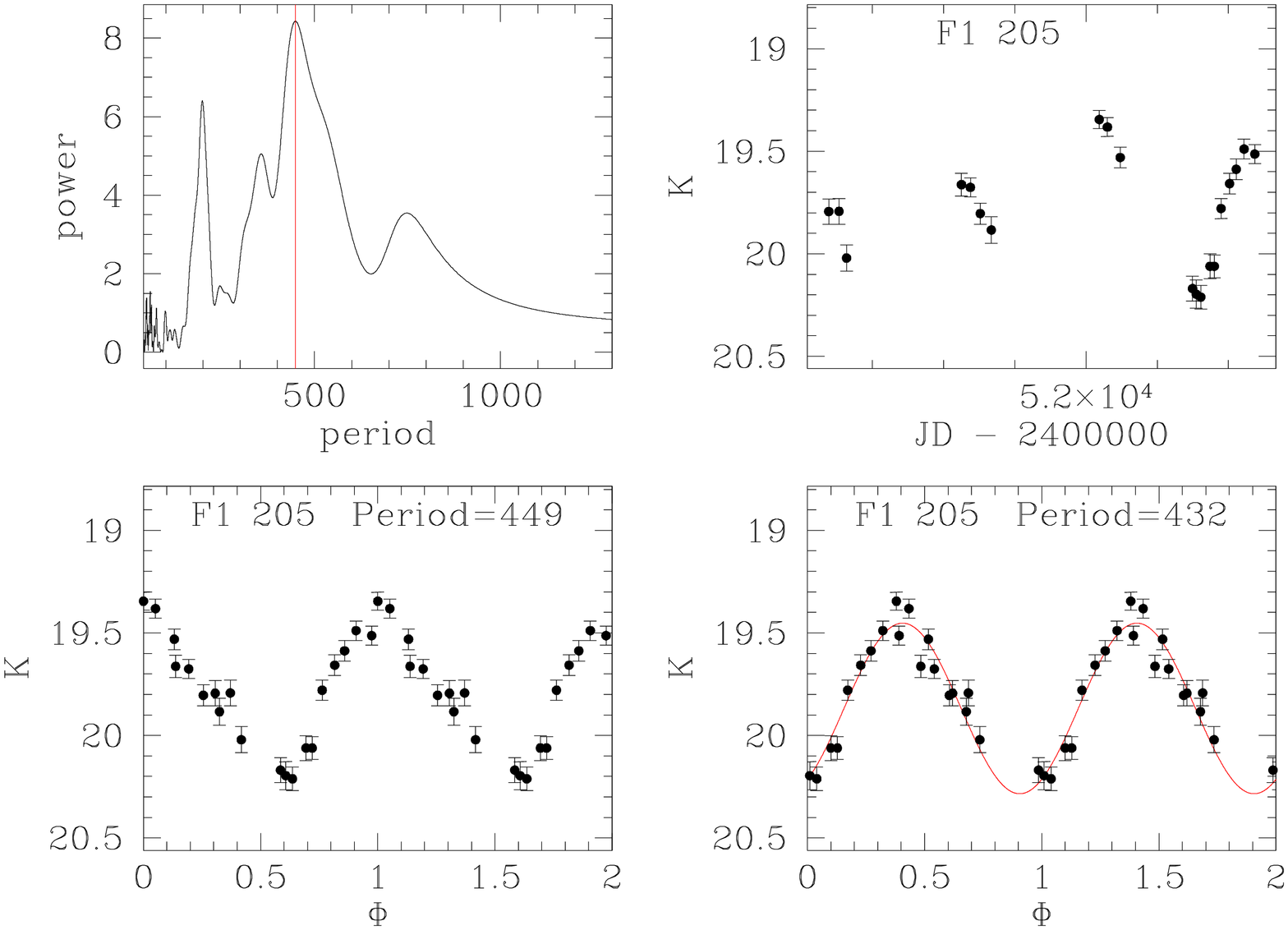}
\includegraphics[width=6.6cm,angle=270]{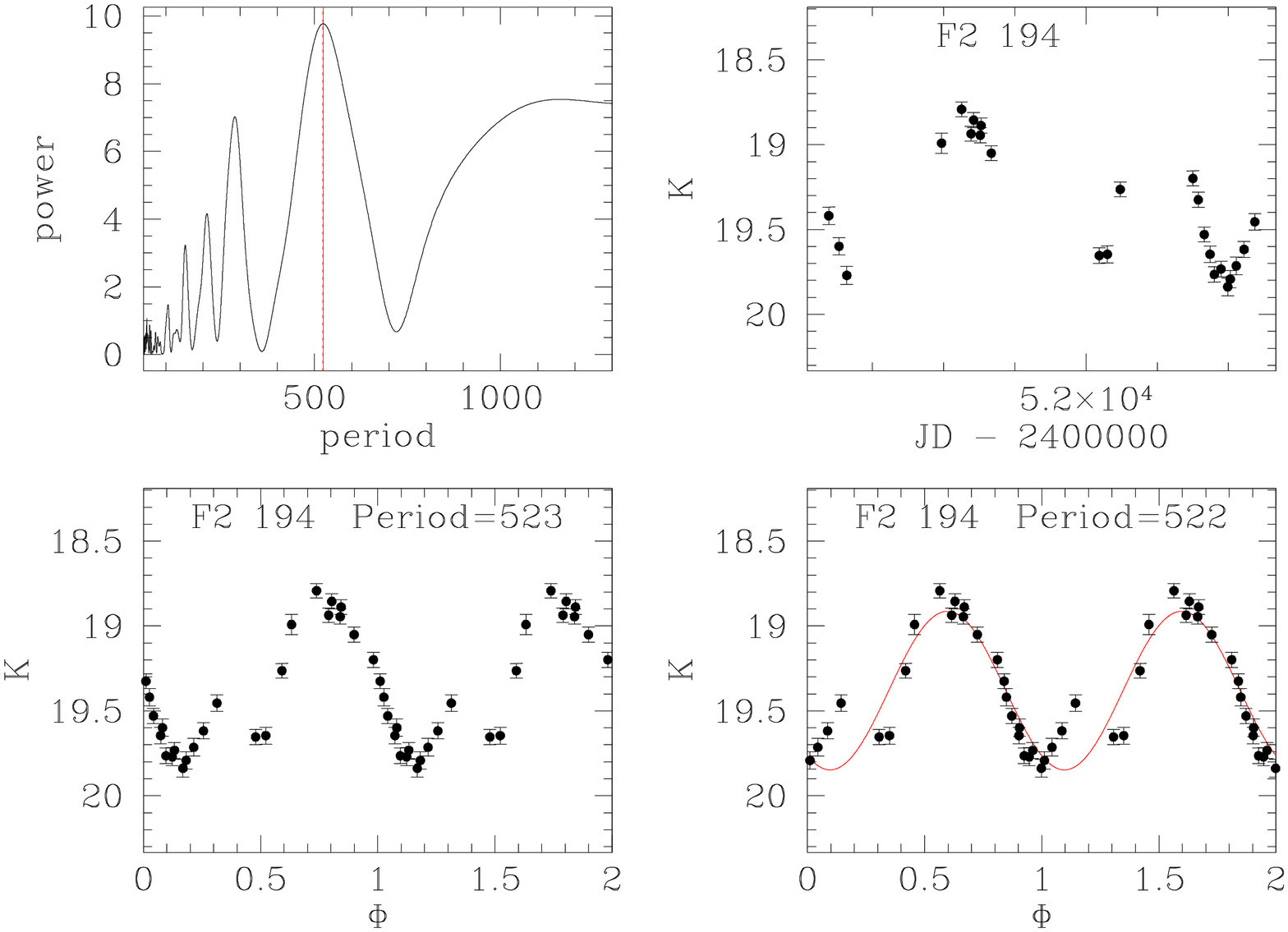}
  \caption[]
	{From left to right, top: periodogram and the
	original light curve in time
	domain; bottom: the light curve folded with the period obtained
	from the highest power frequency from the periodogram 
	and the same light curve folded by the best period obtained from 
	sine curve fitting. Each point in the phased light curves
	is plotted twice to emphasize the periodicity. 
	Over the last light curve the fitting function
	is over-plotted. In the left panels the four diagrams are plotted for
	star \#205 in Field~1 and in the right panels for star \#194 in
	Field~2.}
  \label{P_sinfit}
\end{figure*}

In optical passbands Miras often have
asymmetric light curves, usually steeply rising to the maximum and
with a shallower decline. In near-IR the variations are more regular and
nearly sinusoidal (e.g. Whitelock et al.~\cite{whitelock+00}). Hence a
sine-wave gives a reasonable approximation to most of the LPVs. 

\begin{figure*}
\centering
\includegraphics[width=8.9cm]{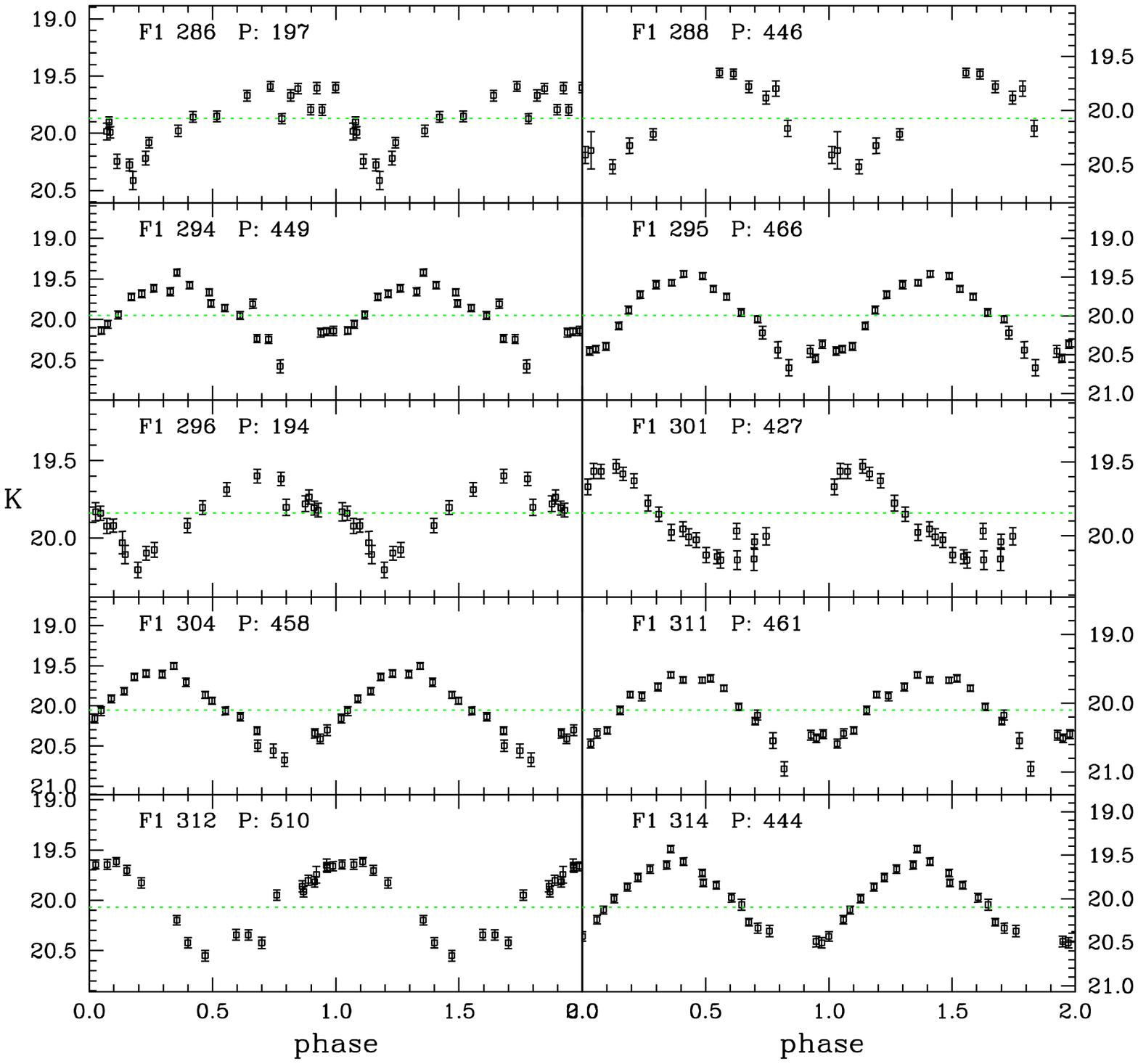}
\includegraphics[width=8.9cm]{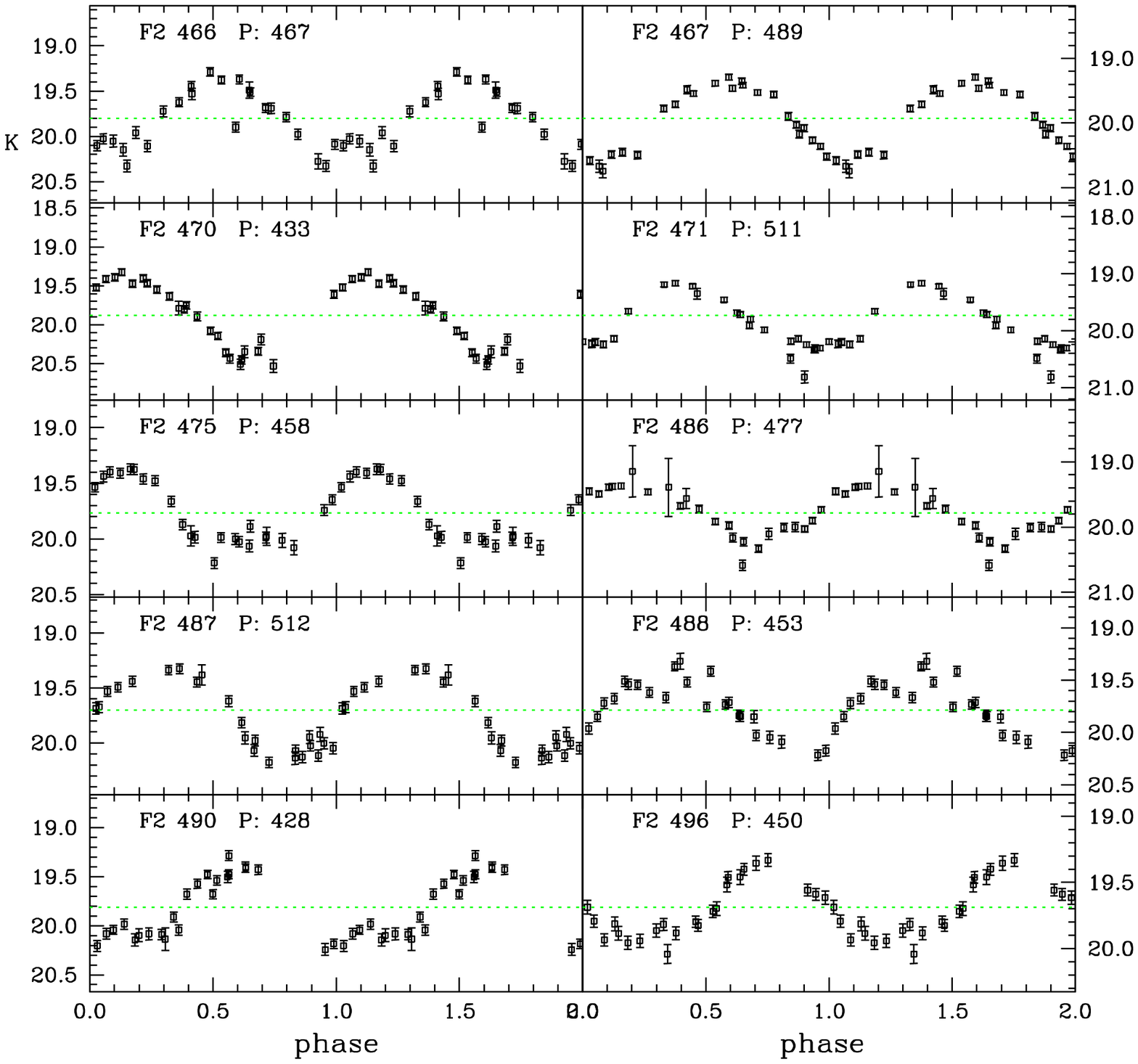}
%
\includegraphics[width=8.9cm]{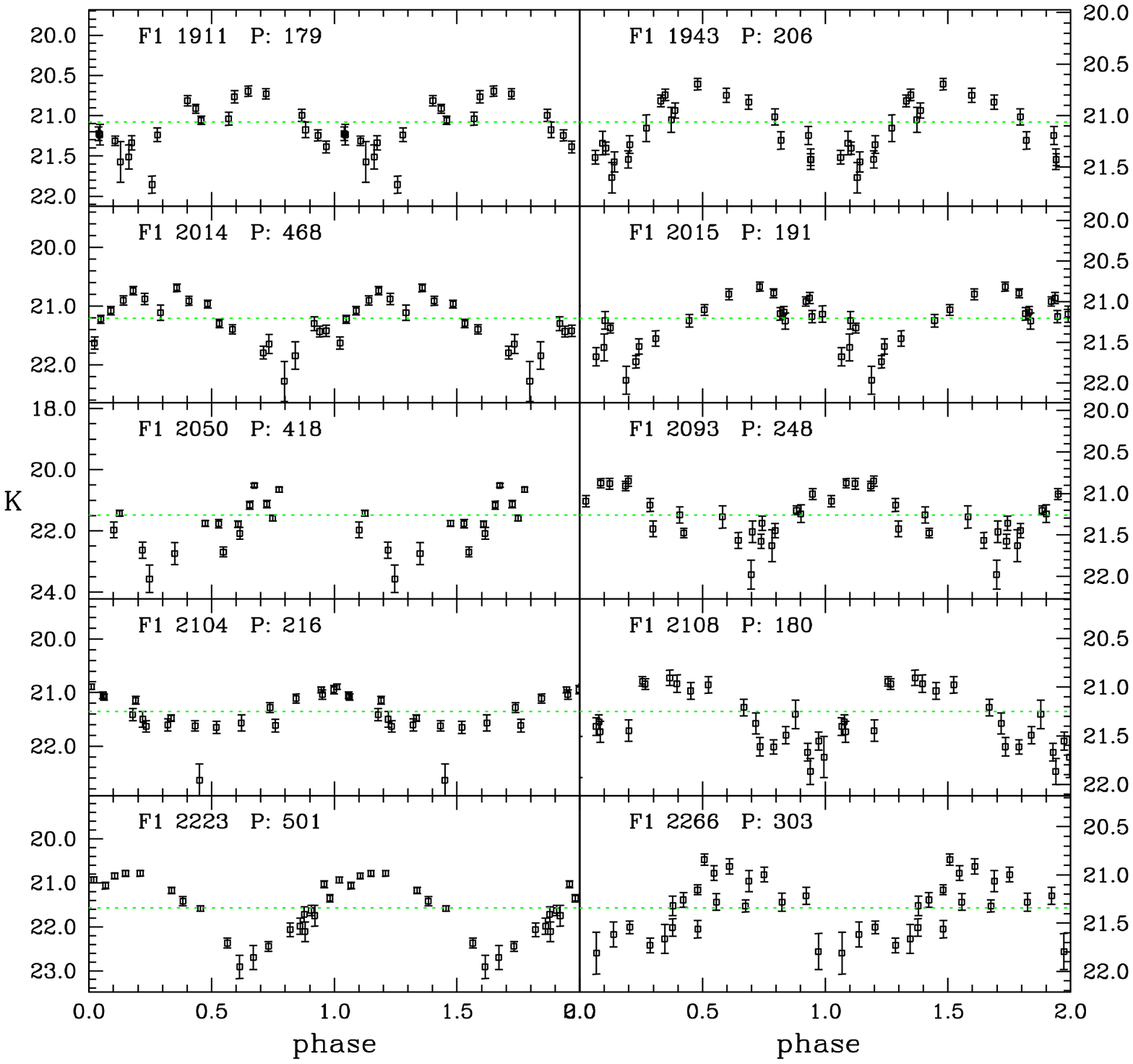}
\includegraphics[width=8.9cm]{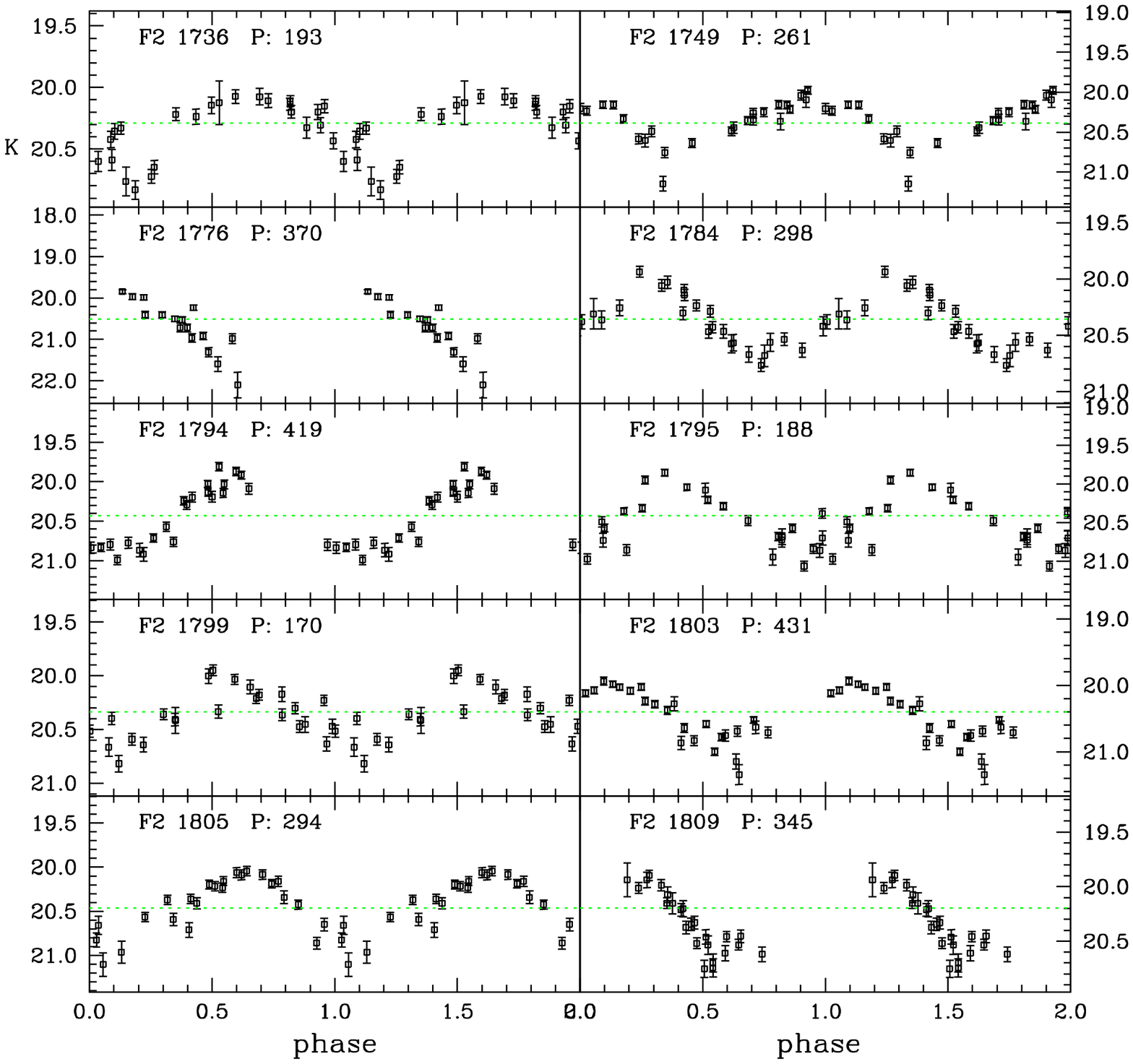}
  \caption[]
        {A sample of phased light curves for variables in both fields and
        with a range of mean magnitudes.
        Each point is plotted twice
        to emphasize the variability. }
 \label{bestPlightcurves}
\end{figure*}

\begin{table*}
\centering
\caption[]
	{Data for a sample of variables with light curves plotted in
	Fig.~\ref{bestPlightcurves}. The columns from left
	to right list: star ID, x and y position with respect to the
	reference epoch, the best fitting period, semi-amplitude, 
	reduced $\chi^2$ of the sine-curve fit, significance of the period
	from Fourier analysis, single epoch $J$ and $H$-band 
	magnitudes, and mean $K$-band magnitude from the sine-curve fit.} 
\begin{tabular}{cccccrrrrrr}
\hline \hline
     ID   &      x    &      y    &  N  &       P &      A    & $\chi^2$ &   signif &        J  &       H   &     $<K>$ \\
\hline
\hline
F1    286 &   279.230 &    67.260 &  20 &     197 &     0.247 &     9.20 &    0.550 &     20.96 &     20.24 &     19.87 \\ 
F1    288 &    14.460 &   549.217 &  11 &     446 &     0.408 &     1.70 &    0.690 &     21.87 &     20.99 &     20.07 \\ 
F1    294 &   124.558 &   330.030 &  20 &     449 &     0.387 &     5.20 &    0.160 &     21.07 &     20.23 &     19.95 \\ 
F1    295 &   445.246 &   622.988 &  20 &     466 &     0.527 &     2.20 &    0.060 &     21.44 &     20.60 &     20.00 \\ 
F1    296 &   395.486 &    18.707 &  19 &     194 &     0.217 &     2.50 &    0.260 &     20.78 &     20.11 &     19.84 \\ 
F1    301 &   364.732 &   570.560 &  20 &     427 &     0.283 &     1.30 &    0.080 &     20.86 &     20.35 &     19.84 \\ 
F1    304 &   169.875 &   226.678 &  20 &     458 &     0.483 &     1.50 &    0.050 &     21.13 &     20.32 &     20.05 \\ 
F1    311 &    51.930 &   187.867 &  20 &     461 &     0.508 &     3.40 &    0.090 &     22.02 &     20.78 &     20.10 \\ 
F1    312 &   326.187 &   501.922 &  20 &     510 &     0.436 &     2.20 &    0.160 &     21.16 &     20.43 &     20.07 \\ 
F1    314 &   108.418 &   332.330 &  20 &     444 &     0.521 &     1.40 &    0.050 &     21.13 &     20.46 &     20.10 \\ 
F1   1911 &   865.938 &   292.732 &  19 &     179 &     0.315 &     5.10 &    0.690 &     22.26 &     21.61 &     21.08 \\ 
F1   1943 &   652.101 &   471.075 &  20 &     206 &     0.366 &     2.10 &    0.310 &     21.61 &     21.31 &     21.06 \\ 
F1   2014 &   862.100 &    33.721 &  19 &     468 &     0.496 &     2.80 &    0.300 &     22.99 &     21.74 &     21.21 \\ 
F1   2015 &   644.197 &   327.874 &  20 &     191 &     0.320 &     4.40 &    0.550 &     21.73 &     21.19 &     21.21 \\ 
F1   2050 &    54.552 &   767.989 &  15 &     418 &     0.821 &    26.00 &    0.960 &     23.04 &     99.99 &     21.48 \\ 
F1   2093 &   719.435 &   768.779 &  20 &     248 &     0.359 &     2.60 &    0.410 &     21.98 &     21.44 &     21.26 \\ 
F1   2104 &   728.300 &   354.414 &  19 &     216 &     0.376 &     2.90 &    0.680 &     24.61 &     23.22 &     21.35 \\ 
F1   2108 &   722.411 &   633.752 &  20 &     180 &     0.053 &    12.90 &    0.380 &     22.59 &     22.11 &     21.25 \\ 
F1   2223 &   760.724 &   516.453 &  20 &     501 &     0.823 &     3.10 &    0.070 &     26.77 &     23.39 &     21.57 \\ 
F1   2266 &   324.743 &   541.049 &  20 &     303 &     0.311 &     6.50 &    0.760 &     21.98 &     21.52 &     21.34 \\ 
F2    466 &   497.668 &   115.838 &  24 &     467 &     0.381 &     7.50 &    0.060 &     21.03 &     20.99 &     19.80 \\ 
F2    467 &   219.420 &   287.679 &  24 &     489 &     0.615 &     3.20 &    0.010 &     22.27 &     21.54 &     19.93 \\ 
F2    470 &   413.810 &   765.776 &  24 &     433 &     0.514 &     2.50 &    0.010 &     20.82 &     20.00 &     19.88 \\ 
F2    471 &   654.271 &   480.820 &  24 &     511 &     0.583 &     3.70 &    0.060 &     22.51 &     21.36 &     19.74 \\ 
F2    475 &   388.723 &   379.739 &  24 &     458 &     0.327 &     4.70 &    0.040 &     20.66 &     20.03 &     19.76 \\ 
F2    486 &    12.504 &   328.661 &  23 &     477 &     0.448 &     3.60 &    0.030 &     20.64 &     20.08 &     19.78 \\ 
F2    487 &   215.930 &   745.931 &  24 &     512 &     0.432 &     2.90 &    0.040 &     21.57 &     20.91 &     19.70 \\ 
F2    488 &   409.287 &   716.276 &  24 &     453 &     0.235 &     9.10 &    0.070 &     20.75 &     20.23 &     19.78 \\ 
F2    490 &   597.413 &    11.477 &  24 &     428 &     0.377 &     5.70 &    0.030 &     21.41 &     20.70 &     19.81 \\ 
F2    496 &   442.349 &   576.357 &  24 &     450 &     0.169 &     6.40 &    0.030 &     21.31 &     20.46 &     19.69 \\ 
F2   1736 &    40.346 &   627.771 &  24 &     193 &     0.247 &     5.20 &    0.240 &     22.44 &     21.37 &     20.29 \\ 
F2   1749 &   832.748 &   178.959 &  24 &     261 &     0.319 &     3.70 &    0.210 &     21.06 &     20.52 &     20.39 \\ 
F2   1776 &   337.907 &    -7.156 &  16 &     370 &     0.607 &    11.10 &    0.660 &     24.63 &     22.32 &     20.51 \\ 
F2   1784 &   711.917 &   559.903 &  24 &     298 &     0.330 &     2.00 &    0.030 &     21.84 &     21.10 &     20.36 \\ 
F2   1794 &   491.605 &    75.748 &  23 &     419 &     0.488 &     4.00 &    0.020 &     22.75 &     21.87 &     20.43 \\ 
%
F2   1795 &   405.954 &   409.103 &  24 &     188 &     0.394 &    10.60 &    0.410 &     21.90 &     20.98 &     20.41 \\ 
F2   1799 &   575.492 &   194.880 &  24 &     170 &     0.231 &     5.80 &    0.240 &     22.02 &     21.10 &     20.33 \\ 
F2   1803 &   786.644 &   606.898 &  24 &     431 &     0.405 &     5.30 &    0.080 &     21.38 &     20.70 &     20.40 \\ 
F2   1805 &   359.442 &   396.001 &  24 &     294 &     0.356 &     4.30 &    0.090 &     21.24 &     20.57 &     20.46 \\ 
F2   1809 &   849.471 &   740.460 &  24 &     345 &     0.386 &     3.30 &    0.060 &     21.56 &     20.96 &     20.19 \\ 
 \hline
\end{tabular}
\label{bestPlightcurves.tab}
\end{table*}

Our code that determines the best fitting period was tested
on a subsample of data on Mira variables published by Whitelock et
al.~(\cite{whitelock+00}). Their data are similar to ours, with 10--15
data points per star and photometric errors of the order of 0.05 mag in K.
In 8 out of 10 cases we obtain the same period as Whitelock et al.
In two cases our periods were significantly different (in one case 
we derived 74 days period with respect to the published value of 185
days and in the second case our period is 422 days where Whitelock
et al. got 327 days), but the light
curves folded with these periods looked as good or even smoother
than the published light curves. In some cases is possible to obtain more
than one acceptable period when there are less than $\sim 15$ 
data points. It
is not clear however, if these stars really pulsate with two different
periods or the insufficient data allow to derive such different periods due
to aliases in the sampling of the data. 

In Figure~\ref{P_sinfit} we show two examples of a
periodogram, the original light curve in time
domain and the light curves folded with the period obtained
from the highest power frequency from the periodogram and by a sine curve
fitting. In a few cases there are evidences for the presence of the second
period or a luminosity modulation
in the data (e.g. star F2 \#194 in Figure~\ref{P_sinfit}), but we
have determined only the main period, because for most of the variables
the data were not good enough for a more detailed analysis.
Most of the times when a good period could be derived, the two periods were
similar, but sometimes they differed by up to 20-30 days and
both light curves were of acceptable quality. This illustrates
that the accuracy of the derived periods and amplitudes varies considerably
from one star to the next, depending on the number
of observations, their individual errors and the stability
of the light curve over the sampling interval. 

It is interesting to note that some LPVs in the LMC with periods in
excess of $\sim 420$ days, which appear to be rather bright for
their periods when compared to other LPVs in the K-band or M$_{\rm bol}$ 
vs. $\log {\mathrm P}$ diagram, show a pronounced hump on the rising branch.
These stars are tentatively identified as being in the Hot Bottom 
Burning stage of evolution on account of, for example, the strong 
Li apparent in some of their spectra 
(Glass \& Lloyd Evans \cite{glass&evans}). Star F2 \#194 shows such a hump
and it is lying 0.28 mag
above the mean period-luminosity relation (Rejkuba 2003, in preparation).
Reviewing the light curves of all the variables with $P>420$~days that are
at least 0.3 mag brighter than the mean K-band magnitude at a given period, 
we found that approximately 1/3 ($\sim 15$ stars) present a similar hump. 
For some of the other 2/3 stars the light curve coverage is not 
sufficiently good to exclude the presence of a hump.

\begin{figure}
\centering
\includegraphics[width=6.55cm,angle=270]{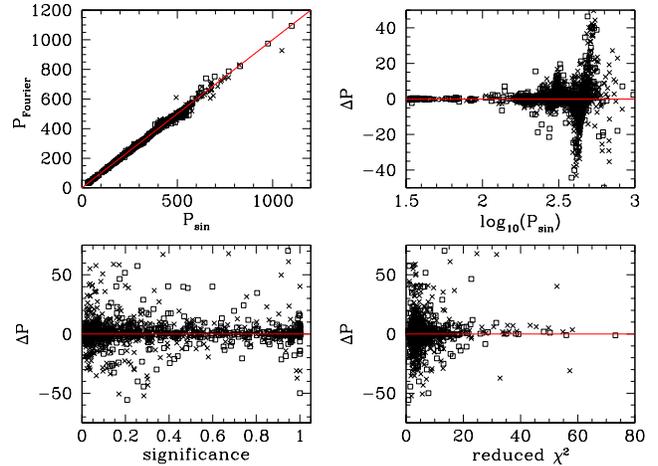}
\caption[]{Comparison of periods derived through Fourier analysis
	(P$_{\rm Fourier}$) and sine-curve fitting 
	(P$_{\rm sin}$) for variables in Field~1 (open squares) 
	and Field~2 (crosses) and the dependence of the differences between
the two methods on quality of the fit parameters: significance measures
strength of the periodicity signal in Fourier analysis and reduced $\chi^2$
is a direct indicator of the quality of the sine-curve fit.}
 \label{compareP}
\end{figure}

The comparison of the periods obtained through Fourier analysis and by a
sine-curve fitting is shown in Figure~\ref{compareP}. Variables in Field~1
are shown with open squares and those in Field~2 with crosses. The upper left
panel is a direct comparison of the periods determined with the two methods,
while the other panels show how the difference depends on period (upper
right), on significance of the period obtained from Fourier analysis (lower
left) and on reduced $\chi^2$ of the sine-curve fit. There is no
significant difference between the two fields. The larger differences are
more probable for variables with low $\chi^2$ of the sine-curve fit. This is
an expected result due to the fact that not all the variables have strictly
sinusoidal light curves. On the contrary, there is no strong dependence on
the significance parameter which measures the strength of the peak in the
periodogram (e.g. Figure~\ref{P_sinfit}). Significance is a smaller number for
stronger (more significant) period. We have visually inspected all the
variables folded with both periods and in cases where one of the two gave
clearly much smoother light curve that period was retained, otherwise the final
period is a mean of the two.
A more quantitative accuracy of the  periods, as well 
as the completeness of the detection of variables with a given 
period was derived using Monte Carlo simulations (see next section).

For 99 variable stars in Field~1 and 169 in Field~2 
no acceptable periods could be obtained because
of the non-sinusoidal variations, large errors combined with small
amplitudes, periodicity outside our probed range ($30 < P \la 1000$ day),
presence of multiple periods, irregularity (cycle-to-cycle variations) 
in Miras or semiregulars, or absence of period (e.g.\ microlensing,
background AGN or SN). More
data are necessary to determine the nature of these variables. Among these
are also LPVs that have periods longer than $\sim1000$ days, 
for which our observations did not cover much more than 1 period and thus
it was not possible to determine an accurate period.

The light curves for 437 variables in Field~1 and 709 in Field~2, for which
we could determine periods, as well as the tables with 
the best fitting parameters
for these stars, are presented only in the electronic version of the article. 
In Figure~\ref{bestPlightcurves} we show a sample of light curves 
folded with the periods that are indicated in each panel. In the example
there are bright and faint variables from both fields.
Table~\ref{bestPlightcurves.tab} lists
for these stars:  ID number, 
x and y position with respect to the reference frame (1Ks07 and 2Ks10 for
Field~1 and Field~2, respectively), number of epochs
with magnitude measurements with $\sigma < 0.5$, periods, semi-amplitudes
($\sqrt{A^2+B^2}$) as
obtained from sine-curve fitting, 
reduced $\chi^2$ of the fits, significances, the $J$- and $H$-band 
magnitudes, and the $K$-band magnitudes from the sine-curve fit.
Negative numbers for x and y positions mean that the star
was out of the limits of the reference frame. The initial 
positions (x,y)=(1,1) correspond to $(\alpha, \delta)= (13^h26^m17\fs5,
-42^{\circ}51\arcmin03\farcs3)$
for Field~1 and to $(\alpha, \delta) = (13^h25^m18\fs0,
-43^{\circ}09\arcmin03\farcs0)$ for Field~2. 
This table is a subset of 
Tables~\ref{JHK1lpv.tab2} and \ref{JHK2lpv.tab2} 
presented in the electronic version.
Additionally, in the electronic version of the article we list all the raw
$K$-band measurements for 1504 variables in both fields
(Tables~\ref{K1all.tab4} and \ref{K2all.tab4}). Magnitude 99.99 and
the error 9.99 denote that no measurement for that epoch was obtained.
Here we give only first three lines of these tables.

\input{H4364T3s.tex}
\input{H4364T4s.tex}
\input{H4364T5s.tex}
\input{H4364T6s.tex}
\input{H4364T7s.tex}
\input{H4364T8s.tex}

It should be noted that while the large majority of the variables with the
determined periods are long period variables with periods in excess of 100
days, there are 27 stars in Field~1
and 13 stars in Field~2 for which most probable periods were shorter than 51
days. However, the significance of these periods is in all cases is very
low, with significance parameter $>0.8$ for all but two stars in Field~1 for
which $\mathrm{signif} >0.65$. If the quoted periods are real, 
these stars could be Cepheids or
some other kind of variables. Cepheids are expected to be found among younger
populations in star forming regions like that present in our Field~1 
(Rejkuba et al.~\cite{rejkuba+01}). Unfortunately our sampling is 
not good enough to determine
reliably the variability type from the light curve shape.

\section{Completeness and contamination}
\label{compl_cont_section}

\begin{figure}
\centering
\includegraphics[width=6.55cm,angle=270]{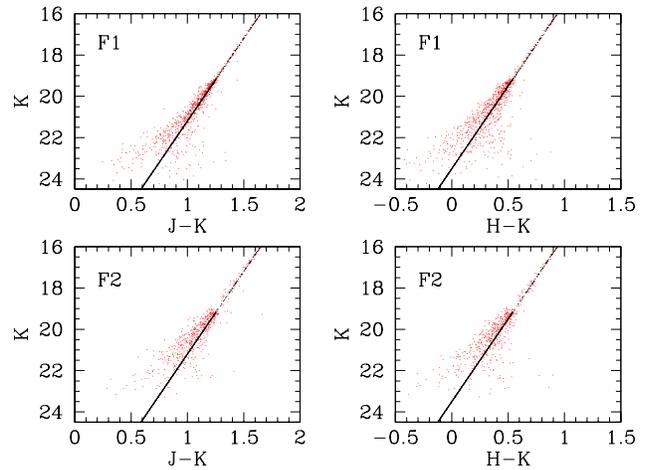}
  \caption[]
	{Completeness simulations: black dots distributed in a line that
	represents a mean RGB ridge line were added to all
	frames. Red dots scattered in the vicinity of that line
	are measured color-magnitude diagrams for these added
	stars. Stars simulated in the Field~1 crowding experiments 
	are shown in the top
	panels and those in Field~2 are in the bottom panels.}
  \label{simcmdsk}
\end{figure}  

The completeness of the photometric catalogue in the 
$J$, $H$ and $K$-bands was
determined with crowding experiments. In ten different crowding
experiments a set of
stars with magnitudes uniformly distributed 
between $15.7<K<25.2$ was added to all the images after the
appropriate re-scaling for the photometric zero point differences and
shifting the stars so that they all fall at the same physical position in
the sky coordinates. The $J-K$ and $H-K$ 
colors of the stars were chosen 
to follow a mean ridge line of the observed red giant branch
(Figure~\ref{simcmdsk}). A realistic,
observed, amount of Poissonian noise was added as well.
The photometry was then re-calculated in the same way as described
above. 

\begin{figure}
\centering
\includegraphics[width=6.55cm,angle=270]{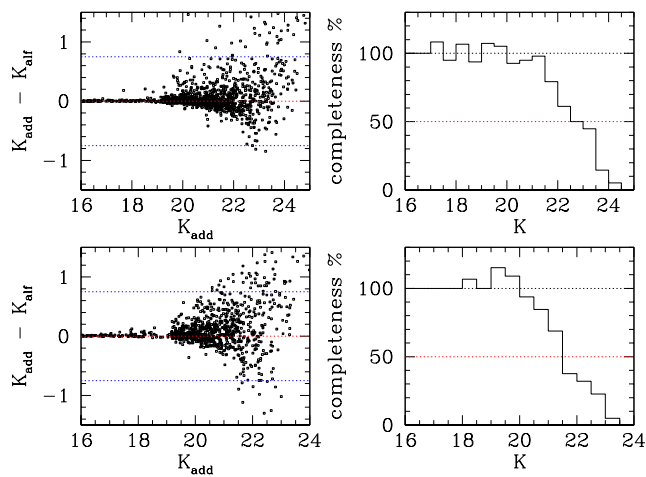}
  \caption[]
	{$K$-band photometric errors (left) and completeness (right):
	The simulations for Field~1 are in the upper and for Field~2 in the
	lower diagrams. Only stars that were measured at the same
	coordinates as the added stars and having $J$, 
	$H$ and $K$-band magnitudes within 
	$-0.75 < \delta K < 0.75$ (dotted lines on the left diagrams) 
	are considered to be detected in the crowding experiments.}
  \label{kcompl}
\end{figure}

\begin{figure}
\centering
\includegraphics[width=6.55cm,angle=270]{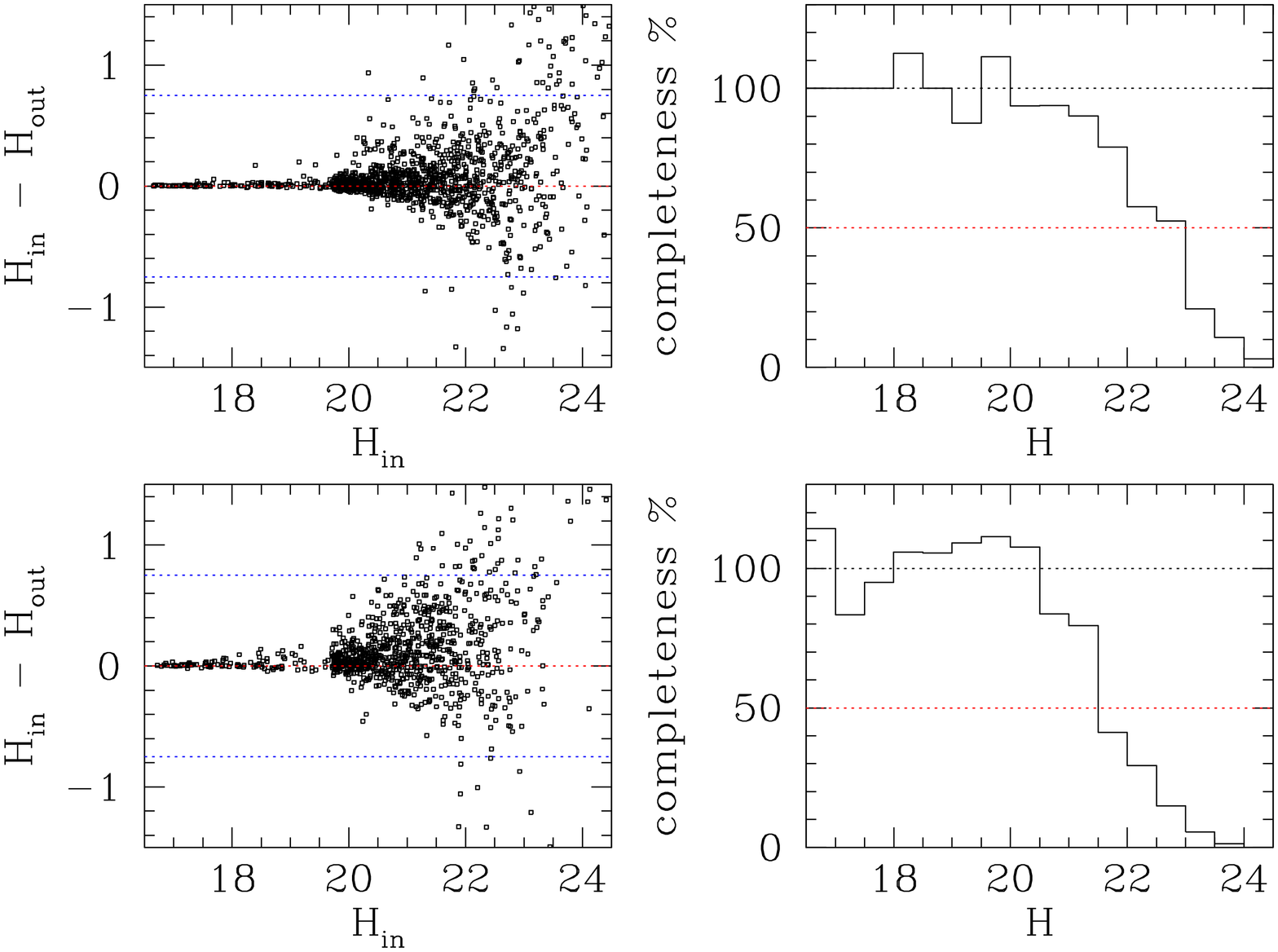}
  \caption[]
	{Same as Fig.~\ref{kcompl} but for the $H$-band.}
  \label{hcompl}
\end{figure}  

\begin{figure}
\centering
\includegraphics[width=6.55cm,angle=270]{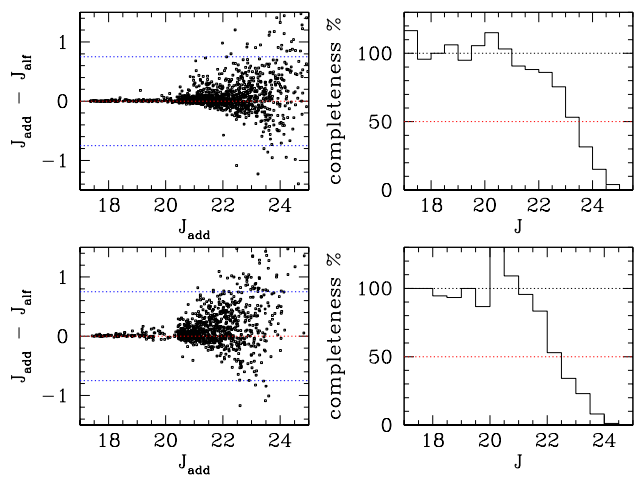}
  \caption[]
	{Same as Fig.~\ref{kcompl} but for the $J$-band.}
  \label{jcompl}
\end{figure}  

The results of these completeness experiments are shown in
Figures~\ref{kcompl}, \ref{hcompl} and \ref{jcompl}. 
Our photometry is complete more than 50\% in the $K$ and $H$-bands for stars 
brighter than 22.5 and 21.5 mag
for fields 1 and 2, respectively. These numbers for the $J$-band are 22.75 and
22.25 mag. Some bins have completeness larger than 100\% due to false
detections or migrations due to blends with original stars in the images.
Note however, that we assume that the simulated star is detected only if its
measured magnitude does not differ from the input value by more than 0.75
mag. 

The left panels of Figures \ref{kcompl} to \ref{jcompl} show differences
between input (added) and recovered (measured) magnitudes which is
indicative of photometric errors at a given magnitude. From them 
it is evident that the stars fainter than 50\% completeness have
very large photometric errors. 
Apart from the completeness and photometric error estimations, these
simulations were also used to determine criteria for the detection of
variable stars. The results are summarized in
Figure~\ref{magNLaddconst} and are described in Section~\ref{varsearch}.

\begin{figure}
\centering
\includegraphics[width=6.55cm,angle=270]{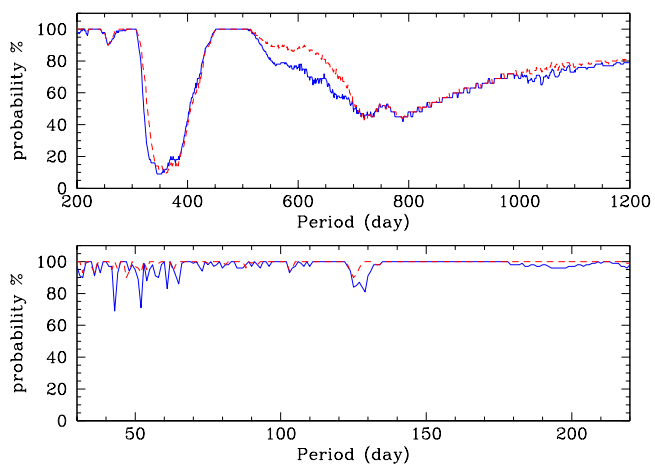}
  \caption[]
	{Probability of detecting a variable star as a function of period
	assuming an even distribution of measurements in phase (see text).
	The solid line is for Field~1 and the 
	dashed line for observations of Field~2.}
  \label{pp}
\end{figure}  

Further simulations of variable stars were used to gain information on the
completeness and contamination in our LPV catalogue. Since the detection
probability of
a variable depends not only on the 
mean magnitude of the star, but also on period
and amplitude, as well as on sampling distribution, 
different sets of simulations were carried out. 

Dependence of the probability of the detection of a variable star with a
given period on the actual distribution of the
observations can be estimated with numerical simulations similar to those by
Saha \& Hoessel~(\cite{saha&hoessel90}) and Bersier \&
Wood~(\cite{bersier&wood02}). Given a period P and initial phase $\phi_0$ at
the time $t_0$ of the first observation, the phase distribution can be
calculated for our set of observations. Assuming that to detect a
variable a uniform coverage in phase is required, with (1) at least two
observations between phases 0 and 0.2, (2) at least two detections with
phases between 0.2 and 0.5 and (3) at least two measurements with phases
between 0.5 and 0.8, the resulting period detection probability is presented
in Figure~\ref{pp}. Note 
that for a more realistic estimate this curve should be multiplied by the
incompleteness corresponding to the mean magnitude of the variable.
An additional factor in the period detection probability 
is the amplitude of the variable.

A better estimate of the detection probability for 
variables with given magnitudes, amplitudes and periods, can be obtained
through simulations similar to the crowding experiments. In a series of
such experiments we have added artificial variable stars with sinusoidal
light curves with given periods, amplitudes and magnitudes. 
The magnitude range
was chosen between $19.5<K<21.5$, similar to that of detected LPVs in NGC
5128 halo. 

\begin{figure}
\centering
\includegraphics[width=6.55cm,angle=270]{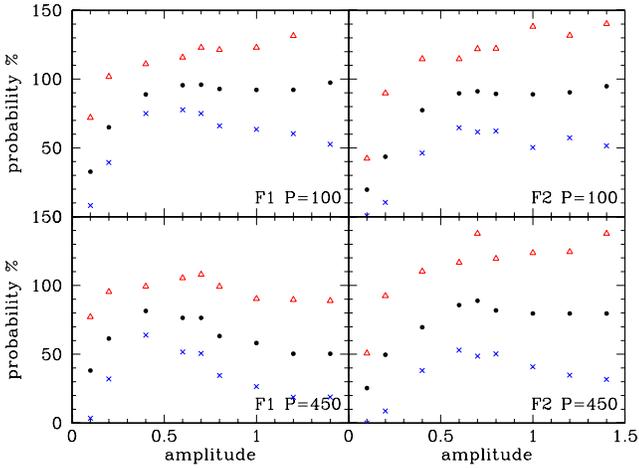}
  \caption[]
	{Probability of detecting a variable star as a function of amplitude
	for simulated variable stars with periods of P=100 day (top panels) and
	P=450 days (bottom panels). A simulated variable is considered to be
	detected only if $|\Delta mag| < 0.75$ and $|\Delta P| \le 25$.
	Simulations for Field~1 are on the 
	left and those for Field~2 on the right. Different symbols are used 
	for different magnitude bins: triangles for stars brighter than K=20.5,
	crosses for $K>20.5$ and filled dots for all the stars irrespective of
	their magnitude.}
  \label{pampl}
\end{figure}  

\begin{figure}
\centering
\includegraphics[width=6.55cm,angle=270]{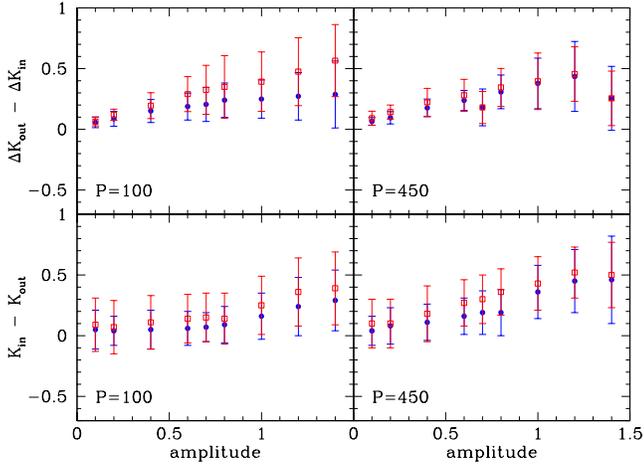}
  \caption[]
	  {The difference between the detected and input amplitude (top panels)
	  and the difference between the detected and input magnitudes 
	  (bottom panels) for simulated variables with periods of P=100 days
	  (left) and P=450 days as a function of input amplitude. 
	  Simulations of Field~1 variables are shown
	  with filled circles and those for Field~2 with open squares.}
  \label{sig_ampl}
\end{figure}  

One series of completeness experiments tested our detection sensitivity to the 
amplitudes of variable stars. Variable stars with amplitudes between 0.1 and
1.4 mag for two fixed periods, 100 and 450 days
were added to the original frames in 18
different crowding experiments for each field.  
In each single experiment all the 
variables had the same amplitude and period, but a range of magnitudes as
mentioned above. In Figure~\ref{pampl} the percentage of detected
variables, those with $-0.75 \le \Delta \mathrm{mag} \le 0.75$ and 
$-25 < \Delta \mathrm{P}<25$, 
as a function of amplitude is given for variables brighter than 
$K=20.5$ (triangles) and those with $K>20.5$ (crosses). Filled dots denote
detection probability (or completeness) as a function of amplitude of variables
for all the stars irrespective of their magnitudes. It is obvious from the fact
that the bright stars have completeness higher than 100\% that there is some
migration of faint stars into the brighter magnitude bin. 

This effect
can also be seen in Figure~\ref{sig_ampl} where we plot differences between
detected and input magnitudes (bottom panels) and differences in detected
and input amplitudes (top panels) as a function of input amplitude
for all the variables. Simulations of Field~1 variables are plotted with filled
dots and those of Field~2 with open squares. 
There is a systematic bias in the sense that we 
detect stars brighter than they are and with somewhat larger amplitudes. 
The magnitude of this bias is proportional
to the amplitude of the variable. Partially this difference may be due to 
the way the mean output magnitude is calculated, but using different 
means still
produces a slight bias. In order to have an estimate of bias independent 
of an a priori knowledge about the period and phase, because 
for some of the variables we could 
not determine periods and some stars in the field 
escaped detection as variable stars, we plot in Figure~\ref{sig_ampl} 
differences in mean magnitudes as they get calculated by DAOMASTER task in 
DAOPHOT.

\begin{figure}
\centering
\includegraphics[width=6.55cm,angle=270]{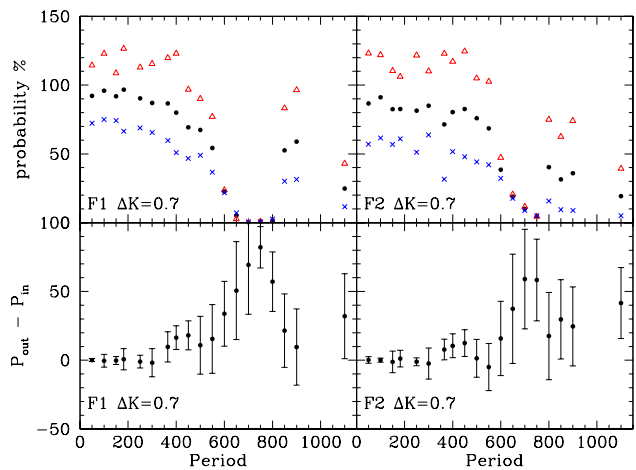}
  \caption[]
	{Probability of detecting a variable star as a function of period
	for simulated variable stars with amplitudes of $\Delta K=0.7$ mag 
	for Field~1 (top left) and Field~2 (top right panel). Here a 
	simulated variable is considered to be detected only if its detected 
	magnitude did not differ from the measured value by more than 0.75 mag
	and if its measured period did not differ from the input one by 
	more than $\pm 25$ days. Bottom panels show difference in measured
	and input periods for variables as a function of period.}
  \label{pper}
\end{figure}  

From Fig.~\ref{pampl} it is obvious that most of the large amplitude 
variables ($\Delta K>0.3$) will be detected for stars with shorter periods.
Our variable star catalogue is highly incomplete for stars with amplitudes
of 0.2 mag or smaller, irrespective of period.
The absolute value of which percentage of variables 
with a given amplitude will be detected for longer period LPVs depends
strongly on our choice of detection criterion. If we consider that a variable
star is detected only if its measured period does not deviate from the input
one by $\pm25$ days, detection probability at P=450 days is between 50 
and 80\%, but if we allow a larger uncertainty in period detection, our 
catalogue is more complete. 

In order to see what is the completeness of our catalogue with respect to 
period of variable stars and how precise our period measurements are 
we performed a series of experiments in which  
we have added to our frames artificial variable stars with sinusoidal
light curves with typical amplitudes for LPVs ($\Delta K=0.7$ mag)
and magnitudes in the range between 
$19.5<K<21.5$, similar to that of detected LPVs in NGC 5128 halo. Periods 
were chosen at ~50 day interval, ranging from 50 to 1100 days. In particular
365 day and 182 day period variable stars were also added. Photometry of 
all the added stars was performed in the same way as before and the light 
curve parameters calculated as explained above. The measured detection
probability as a function of period is presented in Figure~\ref{pper}. Again we
plot the detection probability for all the simulated 
stars with a given period with filled dots. Open triangles and crosses are
used for bright ($K<20.5$) and faint ($K>20.5$) stars, respectively. 
In plotting the top panels of Fig.~\ref{pper} we assumed a variable star
to be detected only if (i) its mean magnitude did not differ from 
the input mean magnitude by more than 0.75 mag, (2) if it was detected 
in at least 10 frames, and (iii) if its measured period did not differ from 
the input period by more than $\pm 25$~days. Due to this last restriction there
is an almost zero probability to detect variables with periods around 700-800
days. However, in our catalogue there are variables with these periods. 
In fact the bottom panels of Figure~\ref{pper} show a dependence of 
the difference between measured and input periods as a function of 
input period. From this it is obvious that for the 
variables with periods in the range between 
700 and 800 days we systematically overestimate their periods by as much as 
50-100 days. In the top panels these variables are considered non-detections.
Periods for variables with periods longer than 800 days can be determined 
more accurately. An
additional potential systematic error could come from period aliasing.
Our observations do not span time continuously. Rather, they are
grouped in three to six month windows with inter-window gaps of 200 to 320
days. Thus, it is possible that stars with actual periods in the range
175 -- 200 days were assigned periods of order of one year, because only
alternate maxima fell within our observing windows.

\section{Conclusions}

We have presented the first catalogue of variable stars in a 
giant elliptical galaxy.
1504 red variables were detected in two halo fields of NGC~5128 (Centaurus
A) covering 10.46 arcmin 
square. For 1146 variables
with at least 10 good measurements we have determined periods, amplitudes and
mean K-band magnitudes using Fourier analysis and non-linear sine-curve 
fitting algorithms. Periods, amplitudes and $JHK$ photometry are given for 
1146 long period variables. Additionally, individual K-band measurements 
are given for all the red variables.

Our extensive completeness simulations show that
periods determined for these long period variables are 
accurate to $\pm25$ days except in the range of 700-800 days where the
accuracy does not exceed $\pm 50$ to 100 days. Additionally there may be 
a few cases where an aliasing period 
may introduce a larger error due to less than optimal sampling.

Mean magnitudes of variable stars
are measured to be brighter than simulated mean magnitudes
by up to $\sim 0.3$ mag and this brightening depends on the amplitude of a 
variable stars. Typically for variables with amplitudes smaller than $\sim 1$ 
mag measured mean magnitudes are very similar to the input magnitudes. This 
brightening is partially due to the way the mean magnitude is calculated.
For constant light curve stars in the same magnitude range as the variable 
stars ($19.5<K<21.5$) there is no bias, i.e. the
difference between measured and input magnitude is 
consistent with zero. 

Our variable star catalogue is close to 90\% complete for short period 
long amplitude variables. For variable stars with variability amplitudes 
of 0.2 mag or smaller detection probability is 50\% or smaller. There is a 
strong dependence of the detection probability on the period of a variable. 
Our simulations show that most variable stars with periods in the range of
700-800 days will have their periods strongly biased toward larger values, 
overestimating the period by as much as 50-100 days.

In the next paper the complete analysis of the period and amplitude 
distribution as well as their dependence on the magnitude and color 
of the variable stars will be presented. The comparison with 
long period variables in the Milky Way and the Magellanic Clouds will
be made.

%
%

\begin{acknowledgements}

We are indebted to many ESO staff 
astronomers who took the data presented in this
paper in service mode operations at Paranal Observatory. Useful input by Tim
Bedding in the early stages of the project is gratefully acknowledged. 
We wish to thank the referee, Tom Lloyd Evans, for useful suggestions
and Ian Glass for communicating their results in advance of the publication.
MR thanks Mariarosa Cioni, Manuela Zoccali and Peter Stetson for
useful discussions. DM is sponsored by FONDAP Center for Astrophysics
15010003. MR acknowledges the ESO studentship programme 
during which most of this work was done.
\end{acknowledgements}

\end{document}

%% file: H4364T3s.tex
\begin{table*}
\centering
\caption[]
	{LPVs in Field 1. The columns from left
	to right list: star ID, x and y position with respect to the
	reference epoch, the best fitting period, semi-amplitude, 
	reduced $\chi^2$ of the sine-curve fit, significance of the period
	from Fourier analysis, single epoch $J$ and $H$-band 
	magnitudes and mean $K$-band magnitude. Magnitude 99.99 
	denotes that no measurement for that star was obtained.} 
\begin{tabular}{cccccrrrrrr}
\hline \hline
 F1  ID &      x    &      y    &  N  &        P &      A    & $\chi^2$ &   signif &        J  &       H   &     $<K>$ \\
\hline \hline
     31 &   460.926 &   401.982 &  20 &     432. &     0.513 &    21.10 &    0.650 &     19.00 &     18.16 &     17.37 \\ 
     53 &   107.785 &   832.967 &  19 &     448. &     0.203 &     2.30 &    0.310 &     19.80 &     19.12 &     18.09 \\ 
     64 &   610.320 &   803.826 &  20 &     430. &     0.202 &    18.50 &    1.000 &     18.92 &     18.57 &     18.36 \\ 
\hline
\end{tabular}
\label{JHK1lpv.tab2}
\end{table*}

%% file: H4364T4s.tex
\begin{table*}
\centering
\caption[]
	{LPVs in Field2. The columns from left
	to right list: star ID, x and y position with respect to the
	reference epoch, the best fitting period, semi-amplitude, 
	reduced $\chi^2$ of the sine-curve fit, significance of the period
	from Fourier analysis, single epoch $J$ and $H$-band 
	magnitudes and mean $K$-band magnitude.Magnitude 99.99 
        denotes that no measurement for that star was obtained.} 
\begin{tabular}{cccccrrrrrr}
\hline \hline
F2   ID &      x    &      y    &  N  &      P   &      A    & $\chi^2$ &   signif &        J  &       H   &     $<K>$ \\
\hline \hline
     35 &   115.155 &   -29.046 &  10 &      35. &     0.978 &   226.70 &    0.940 &     18.65 &     18.42 &     18.48 \\ 
     49 &    41.708 &   172.990 &  10 &     988. &     1.463 &     2.20 &    0.980 &     99.99 &     99.99 &     17.87 \\ 
     61 &   591.542 &   445.812 &  24 &     696. &     0.180 &     1.70 &    0.300 &     19.56 &     18.77 &     18.41 \\ 
\hline
\end{tabular}
\label{JHK2lpv.tab2}
\end{table*}

%% file: H4364T5s.tex
\begin{table*}
\centering
\caption[]
	{Additional variable stars in Field 1 for which periods 
	could not be determined. The columns from left
	to right list: star ID, x and y position with respect to the
	reference epoch,  single epoch $J$ and $H$-band 
	magnitudes and mean $K$-band magnitude. Magnitude 99.99 
	denotes that no measurement for that star was obtained.} 
\begin{tabular}{cccrrrr}
\hline \hline
 F1  ID &      x    &      y    &  N  &       J  &       H   &     $<K>$ \\
\hline \hline
     50 &   763.941 &   462.473 &  20 &     19.59 &     18.87 &     18.23 \\ 
     51 &   219.227 &    79.584 &  20 &     18.37 &     18.13 &     17.97 \\ 
     69 &   113.406 &   403.575 &  12 &     99.99 &     99.99 &     21.80 \\ 
\hline
\end{tabular}
\label{JHK1var.tab3}
\end{table*}

%% file: H4364T6s.tex
\begin{table*}
\centering
\caption[]
	{Additional variable stars in Field 2 for which 
	periods could not be determined. The columns from left
	to right list: star ID, x and y position with respect to the
	reference epoch,  single epoch $J$ and $H$-band 
	magnitudes and mean $K$-band magnitude. Magnitude 99.99 
	denotes that no measurement for that star was obtained.} 
\begin{tabular}{cccrrrr}
\hline \hline
 F2  ID &      x    &      y    &  N  &       J  &       H   &     $<K>$ \\
\hline \hline
      5 &   110.132 &   451.042 &  17 &     99.99 &     18.79 &     15.51 \\ 
     25 &   111.583 &   -16.658 &  14 &     18.85 &     19.00 &     18.39 \\ 
     63 &   133.142 &   -27.505 &   9 &     19.02 &     18.60 &     18.96 \\ 
\hline
\end{tabular}
\label{JHK2var.tab3}
\end{table*}

%% file: H4364T7s.tex
\begin{table*}
\centering
\caption[]
	{Individual measurements of the K-band magnitudes for all the
	red variable stars detected in Field~1. The columns from left
	to right list: star ID, x and y position with respect to the
	reference epoch, the reference epoch K-band 
	magnitudes and errors and all the other individual K-band magnitudes
	and the errors of the measurements as given
	by ALLFRAME to which 0.04 mag calibration error have been added in
	quadrature. Magnitude 99.99 and the
	error 9.99 denote that no measurement for that star was obtained.} 
\begin{tabular}{rrrrrrrrrr}
\hline \hline
 F1  ID &      x    &      y    & 1Ks07 ($\sigma$)&1Ks01 ($\sigma$)&1Ks02 ($\sigma$)&1Ks03 ($\sigma$)&1Ks04 ($\sigma$)&1Ks05 ($\sigma$)&1Ks06 ($\sigma$)\\
&&&1Ks08 ($\sigma$)&1Ks09 ($\sigma$)&1Ks10 ($\sigma$)&1Ks11 ($\sigma$)&1Ks12 ($\sigma$)&1Ks13 ($\sigma$)&1Ks14 ($\sigma$)\\
&&&1Ks15 ($\sigma$)&1Ks16 ($\sigma$)&1Ks17 ($\sigma$)&1Ks18 ($\sigma$)&1Ks19 ($\sigma$)&1Ks20 ($\sigma$)\\
\hline \hline
     31 &   460.926 &   401.982 &     17.38 (0.09) &     17.41 (0.08) &     16.54 (0.08) &     16.70 (0.06) &     17.24 (0.09) &     17.24 (0.06) &     16.48 (0.07) \\
        &           &           &     17.37 (0.05) &     17.50 (0.06) &     17.54 (0.06) &     17.49 (0.05) &     17.26 (0.05) &     17.85 (0.05) &     18.04 (0.08) \\
        &           &           &     18.04 (0.08) &     17.86 (0.09) &     17.88 (0.09) &     17.66 (0.05) &     18.33 (0.11) &     17.67 (0.07) & \\ 
     50 &   763.941 &   462.473 &     18.00 (0.08) &     18.12 (0.05) &     17.98 (0.05) &     17.94 (0.04) &     17.92 (0.07) &     18.00 (0.05) &     17.80 (0.05) \\
        &           &           &     17.85 (0.04) &     18.04 (0.05) &     18.03 (0.07) &     18.09 (0.04) &     17.99 (0.05) &     18.08 (0.05) &     18.19 (0.05) \\
        &           &           &     18.18 (0.05) &     18.18 (0.04) &     18.17 (0.06) &     18.14 (0.05) &     18.21 (0.04) &     18.14 (0.05) & \\ 
     51 &   219.227 &    79.584 &     17.91 (0.04) &     17.83 (0.04) &     17.71 (0.06) &     17.78 (0.05) &     17.71 (0.05) &     17.90 (0.04) &     17.76 (0.05) \\
        &           &           &     18.12 (0.04) &     18.06 (0.04) &     18.07 (0.04) &     18.23 (0.04) &     18.18 (0.04) &     18.24 (0.04) &     18.29 (0.04) \\
        &           &           &     18.23 (0.05) &     18.18 (0.04) &     18.18 (0.05) &     18.06 (0.04) &     18.25 (0.05) &     18.13 (0.04) & \\  
\hline
\end{tabular}
\label{K1all.tab4}
\end{table*}

%% file: H4364T8s.tex
\begin{table*}
\centering
\caption[]
	{Individual measurements of the K-band magnitudes for all the
	red variable stars detected in Field~2. The columns from left
	to right list: star ID, x and y position with respect to the
	reference epoch, the reference epoch K-band 
	magnitudes and errors and all the other individual K-band magnitudes
	and the errors of the measurements as given
	by ALLFRAME to which 0.04 mag calibration error have been added in
	quadrature. Magnitude 99.99 and the
	error 9.99 denote that no measurement for that star was obtained.} 
\begin{tabular}{rrrrrrrrr}
\hline \hline
 F2  ID &      x    &      y    & 2Ks10 ($\sigma$) & 2Ks01 ($\sigma$) & 2Ks02 ($\sigma$) & 2Ks03 ($\sigma$) & 2Ks05 ($\sigma$) & 2Ks06 ($\sigma$) \\
        &           &           & 2Ks07 ($\sigma$) & 2Ks08 ($\sigma$) & 2Ks09 ($\sigma$) & 2Ks11 ($\sigma$) & 2Ks12 ($\sigma$) & 2Ks13 ($\sigma$) \\
        &           &           & 2Ks14 ($\sigma$) & 2Ks15 ($\sigma$) & 2Ks16 ($\sigma$) & 2Ks17 ($\sigma$) & 2Ks18 ($\sigma$) & 2Ks19 ($\sigma$) \\ 
        &           &           & 2Ks20 ($\sigma$) & 2Ks21 ($\sigma$) & 2Ks22 ($\sigma$) & 2Ks23 ($\sigma$) & 2Ks24 ($\sigma$) & 2Ks04 ($\sigma$) \\
\hline \hline
      5 &   110.132 &   451.042 &     14.76 (0.04) &     15.33 (0.14) &     15.04 (0.15) &     15.00 (0.12) &     14.66 (0.05) &     14.69 (0.04) \\ 
        &           &           &     14.68 (0.04) &     14.68 (0.04) &     14.69 (0.04) &     15.37 (0.06) &     15.41 (0.07) &     15.50 (0.13) \\
        &           &           &     16.69 (0.13) &     17.02 (0.16) &     15.97 (0.28) &     16.08 (0.18) &     99.99 (9.99) &     99.99 (9.99) \\
        &           &           &     99.99 (9.99) &     99.99 (9.99) &     99.99 (9.99) &     99.99 (9.99) &     99.99 (9.99) &     14.73 (0.07) \\ 
     25 &   111.583 &   -16.658 &     99.99 (9.99) &     22.55 (0.25) &     99.99 (9.99) &     99.99 (9.99) &     99.99 (9.99) &     99.99 (9.99) \\
        &           &           &     99.99 (9.99) &     99.99 (9.99) &     99.99 (9.99) &     17.57 (0.07) &     18.26 (0.11) &     16.74 (0.04) \\
        &           &           &     18.18 (0.21) &     17.87 (0.11) &     99.99 (9.99) &     99.99 (9.99) &     21.48 (0.34) &     19.79 (0.24) \\
        &           &           &     20.17 (0.09) &     21.46 (0.18) &     18.18 (0.16) &     19.47 (0.09) &     19.93 (0.10) &     19.98 (0.22) \\  
     35 &   115.155 &   -29.046 &     99.99 (9.99) &     99.99 (9.99) &     99.99 (9.99) &     99.99 (9.99) &     99.99 (9.99) &     99.99 (9.99) \\
        &           &           &     99.99 (9.99) &     99.99 (9.99) &     99.99 (9.99) &     17.02 (0.06) &     99.99 (9.99) &     16.71 (0.06) \\
        &           &           &     19.02 (0.27) &     17.75 (0.18) &     24.37 (6.32) &     99.99 (9.99) &     22.20 (0.49) &     99.99 (9.99) \\
        &           &           &     20.37 (0.10) &     21.65 (0.31) &     18.49 (0.19) &     99.99 (9.99) &     21.58 (0.34) &     19.99 (0.23) \\  
\hline
\end{tabular}
\label{K2all.tab4}
\end{table*}